# Application of the X-ray laser to muon-catalyzed *d-t* fusion

Sachie Kimura*, Aldo Bonasera*,#

*INFN Laboratori Nazionali del Sud, Catania, Italy
#Libera Università Kore, Enna, Italy

**Abstract**
We discuss the alpha-muon sticking coefficient in the muon-catalysed d-t fusion in the framework of the Constrained Molecular Dynamics model. Especially the influence of muonic chaotic dynamics on the sticking coefficient is brought into focus. The chaotic motion of the muon affects not only the fusion cross section but also the $\mu$-$\alpha$ sticking coefficient. Chaotic systems lead to larger enhancements with respect to regular systems because of the reduction of the tunneling region. Moreover they give smaller sticking probabilities than those of regular events. By utilizing a characteristic of the chaotic dynamics one can avoid losing the muon in the $\mu$CF cycle. We propose that the application of the so-called "microwave ionization of a Rydberg atom" to the present case could lead to the enhancement of the reactivation process by using X-rays.

## 1. Introduction

The muon catalyzed fusion ($\mu$CF) of hydrogen isotopes, especially *d-t* fusion, has been studied as a realizable candidate of an energy source at thermal energies. In the liquid $D_2$ and $T_2$ mixture, the muon assists the fusion through the formation of a muonic molecule, since the size of the muonic molecule is much smaller than that of the ordinary molecules and the fusing nuclei tend to stay closer. After the fusion process the muon is released normally and again it is utilized for a subsequent fusion until the muon decays. The efficiency of the $\mu$CF is governed by not only the life time of muons but also the muon-sticking on the $\alpha$ particle which is produced in the fusion[1,2]. The muon is lost from the $\mu$CF cycle by the initial sticking($w_0$), unless it is not released through the interaction with the medium. The rate of the stripping of the stuck muon from the $\alpha$ particle is known as the reactivation coefficient $R$ and thus the effective sticking probability ($w_0^{eff}$) is determined by $w_s^{eff} = w_0(1-R)$. In this paper we do not take into account the medium effects which are supposed to be important to determine the precise value of $R$ in the actual experimental setup. We rather aim to propose a method to enhance the reactivation process, by making use of the stochastic instability of the stuck muon in an oscillating field. For this purpose we are mainly interested in investigating the impact of the regular and chaotic dynamics [3]. At thermal energies, where the $\mu$CF takes place, fluctuations are anticipated to play a substantial role. We investigate the influence of the fluctuations by using a semi-classical method, the constrained molecular dynamics (CoMD) approach. In the CoMD, the constraints restrict the phase space configuration of the muon to fulfill the Heisenberg uncertainty principle. The results are given as an average and a variance over ensembles of the quantity of interest which is determined in the simulation. Especially we determine the enhancement factor of the reaction cross section by the muon as a function of the incident energy. The enhancement factor of each event indicates the regularity of the system. Subsequently we determine the initial muon sticking probability, using the phase space distribution of the muon at the internal classical turning point. The muon does not stick necessarily to the ground state of the alpha particle and this fact plays an important role when we proceed to the stripping of the bound muon in the oscillating field. The chaotic dynamics could prevent the muon from being lost in the $\mu$CF cycle due to the sticking. This is achieved by utilizing the characteristic as a nonlinear oscillator of the trapped muon on the alpha particle. We draw an analogy between the muonic He ion in the present case and microwave-ionization of Rydberg atoms [4,5], where the driven electron in a highly excited hydrogen atom in a strong microwave electric field exhibits the chaotic dynamics and is ionized. We carry out a numerical simulation by enforcing an oscillating field (linearly polarized, oscillatory electric field) on the system. This can be, experimentally, achieved by X-ray lasers. The oscillating force causes the resonance between the force itself and the oscillating motion of the muon around the alpha particle, especially when the driving frequency coincides with integer multiples of the eigen frequency of the muonic helium.

## 2. Framework

The detail of the framework of the CoMD is discussed in the references [6,7]. In the following we discuss the application of the CoMD to the simulation of the fusion and the subsequent muon sticking processes.

## 3. Enhancements of the cross section by the muonic screening effect

We introduce the enhancement factor of the cross section by the bound muon $f_\mu$

$$f_\mu = \sigma(E)/\sigma_0(E), \quad (1)$$

where $\sigma(E)$ and $\sigma_0(E)$ are the real cross section and the cross section in the absence of the muon, respectively. In the following discussion, the enhancement factor is referred as an indicator of the regularity of the muonic motion. It plays a role of a sort of order parameter and is determined through the obtained values in the numerical simulation.

[Insert figures 1 and 2 about here]

In figure 1 we plot $f_\mu$ as a function of the incident center-of-mass energy between the triton and the deuteron. From our simulation of the collisions using an ensemble of events, we determine the average enhancement factor $\bar{f}_\mu$ and its variance : $\Delta f_\mu$. These are shown by squares and error-bars, respectively. Both the average enhancement factor and its variance increase exponentially as the incident energy decreases. The dashed line in the figure corresponds to the enhancement factor in the adiabatic limit. The obtained average enhancement factor is in agreement with the adiabatic limit. The arrows in the figure denote the ionization energy of the muonic tritium. At this energy the total energy of the system is zero. At the incident energies higher than this point, the total system is unbound, while the 3-body system is bound at lower energies. In the low energy limit the variance, which is shown by error-bars, becomes large. It implies that the system exhibits a sensitive dependence of the dynamics on initial conditions, i.e. the occurrence of chaos. We indeed verify the manifestation of chaos by plotting the Poincare surface of section with respect to the enhancement factor for two events in figure 2. In the figure we show the surface of section for two selected events at the incident energy 0.18 keV on the $x - p_x$ plane (figure 2 left panels) and on the $z - p_z$ plane (figure 2 right panels), respectively. We choose the beam axis to coincide with the z-axis. At the incident energy 0.18 keV the average enhancement factor, $\bar{f}_\mu = 2.9 \times 10^{29}$ as one can see in figure 1. In the top panels, with 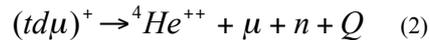 $f_\mu = 4.1 \times 10^{19} (<< \bar{f}_\mu)$ and the ratio of the external classical turning point in the presence of the muon to the one in the absence of the muon: $ctp_\mu / ctp_0 = 0.15$, the points show a map of a typical regular event. By contrast in the bottom panels, with 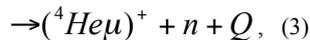 $f_\mu = 2.7 \times 10^{31} (> \bar{f}_\mu)$ and $ctp_\mu / ctp_0 = 0.06$, the points show the map of an irregular event; the points cover a large section of the map. This comparison between two cases suggests that the irregular muonic motion leads to smaller external classical turning point with respect to the regular motion. As a consequence the fusion reaction with the irregular muonic motion has larger enhancements factors.

## 4. Muon sticking probability

We estimate the sticking probability of muons on the alpha particle in the exit channel;

$$(td\mu)^+ \to {}^4He^{++} + \mu + n + Q \quad (2)$$
$$\to ({}^4He\mu)^+ + n + Q, \quad (3)$$

where $Q = 17.59$ (MeV) is the decay Q-value of this reaction. The muon remains bound (3), if the binding energy of the muon on an alpha particle is negative, in the center-of-mass system of the muon and the alpha particle. From this condition we deduce the following equation for the angle $b$ between the velocities of the muon and the alpha particle: $v_\mu$ and $v_a$.

$$\cos b \geq \frac{M_{a\mu}/2 \,(|v_\mu|^2 + |v_a|^2) - 2e^2/|r_{a\mu}|}{M_{a\mu}|v_\mu||v_a|} \qquad (4)$$

Here $M_{a\mu}$ denotes the reduced mass of the muon and the alpha particle. The condition Eq. (4) is fulfilled, when the r.h.s of the equation, let us call it as $g$, is equal to 1 or less and for the solid angle $\Omega = 2pi(1-g)$ [ steradian] in the 3-dimensional space. We can therefore estimate the sticking probability by $\Omega/4pi$, if $g \leq 1$. By inspecting each event, one could find out some events which satisfy the condition $g < 1$. Indeed one of the two events shown in the top panels in figure 2, which is regular, has $g = 0.93(< 1.0)$ therefore the sticking probability of this event itself is not zero (3.8 %), while the other in the bottom panels has $g = 1.9(> 1.0)$ and its sticking probability is zero. In the same way we calculated $g$ for all the events which are created in our simulation. The obtained sticking probability is shown in the left panel in figure 3 as a function of the incident energy with filled circles. [Insert figure 3 about here] At the same time, we carry out the simulation of the exit channel by creating 20000 events (randomly chosen directions of outgoing particles). We distinguish the muon sticking event (3) from the release event (2) by monitoring the binding energy of the muon on alpha particles and the radius of the muonic ion. We count the events where the binding energy of the muon on the alpha particle and the distance between the muon and the alpha particle maintain to be negative and small respectively, up to the point where the alpha is distant enough from the neutron. In the right panel in Figure 3 typical trajectories of the distance between the muon and the alpha particle are shown as a function of the inter-nuclear separation. Among the 6 curves shown in the right panel in the figure, 3 curves, which show oscillational behaviors, are corresponding to the sticking events. While 3 other curves increase monotonically after $R_{n\alpha}$ exceeds 0.01 Å. The horizontal straight line in the figure indicates the size of the radius of the ground state muonic He atom. The obtained sticking probability is shown with open circles with error-bars in the left panel in figure 3. First, as one can see clearly, the result of numerical simulation agrees with the sticking probability which is calculated considering the solid angle. In the figure the solid line with an error bar denotes the sticking coefficient obtained from the direct measurement of the initial sticking probability [2]. The resulting sticking probability ranges nearby the experimental value as a function of the incident energy, except for the several points that have zero and relatively large sticking probability. We mention specially that the incident energy, at which the sticking probability becomes zero, is slightly below than the ionization energy of the muonic tritium.

## 5. Muon stripping

The stuck muon is possibly stripped from the alpha particle, by enforcing a linearly polarized oscillatory field on the system. The periodic motion of the stuck muon can be expressed in terms of nonlinear oscillations. For a nonlinear oscillator the oscillating driving force, i.e., the oscillating field causes a resonance between the force itself and the oscillating motion of the muon at driving frequencies, which are integer multiples of the fundamental frequencies of the muon. In our present case the muonic He is not stuck necessarily in its ground state. So that the muon can be ionized directly from one of such an excited state by radiating the electrostatic wave, otherwise the muon is, at first, prompted to an excited state and then ionized. In either case it will be achieved by using X-ray lasers, since the fundamental frequencies for the ground and the first excited states of the muonic helium ion correspond to 0.11nm and 0.44nm, respectively, in terms of the wave length. In our numerical simulation, instead of the discontinuous frequencies, we obtain the proper frequency of the stuck muon for each event. We simulate the stripping with the above-mentioned external force.
[Insert figure 4 about here]
In figure 4 we show the time T map of the oscillational motion of the muon on the $r$-$p$ plane, where $r$ is relative distance between the muon and the alpha particle and $p$ is its conjugate momentum. We follow 1000 cycles of the driven oscillation in our simulation. We choose two sticking events, which are shown in the bottom panel in Figure 3. The one has smaller amplitude, a tightly bound state (top panels), and the other has larger amplitude, a loosely bound state (bottom panels). The left panels show the map of the stuck muon without external force. The map remains in a limited manifold around $r = 0$. The right panels show the case with the external force $zF\sin(\gamma t)$ with the driven frequency $\gamma = 2\gamma_\mu$, where $\gamma_\mu$ is the angular frequency of the muon in the muonic helium ion. In the case of the tightly bound muon,

$\gamma_\mu = 2.37 \times 10^{18} (s^{-1})$) and the muonic atom is excited in a loosely bound state and then ionized. While in the case of the loosely bound muon, $\gamma_\mu = 2.8 \times 10^{17} (s^{-1})$ and the muonic atom is ionized in one step. The wave lengths of the laser fields, which are used in our simulation, are 0.063 and 0.52 (nm), respectively. The muonic atom in the external oscillating field is captured into a resonant state and ionized due to its stochastic instability. One can see clearly that the muon is expelled from the Helium with the external oscillational force with the corresponding frequency of the unperturbed system. We point out that the muonic molecule is not destroyed by the external force with same frequency which we used in the above discussion.

## 6. Conclusions

We discussed the alpha-muon sticking coefficient in the muon-catalyzed d-t fusion. We performed numerical simulations by using the Constrained Molecular Dynamics model. Especially the influence of the muonic chaotic dynamics on the sticking coefficient is brought into focus. The chaotic motion of the muon affects not only the fusion cross section but also the sticking coefficient. The irregular (chaotic) dynamics of the bound muon lead to larger enhancements with respect to regular systems because of the reduction of the tunneling region. Moreover they give smaller sticking probabilities than those of regular events. We proposed a method to strip the stuck muon from the alpha particle by exposing the system in the X-ray radiation field. Its numerical experiment has been performed under an oscillating external force with the driving frequency twice as high as the angular frequency of the stuck muon. We have observed that the muon has been released successfully with the selected frequency. By utilizing the chaotic dynamics one can prevent the muon from losing in the $\mu$ CF cycle by the sticking.

Figure captions

Figure 1: Enhancement factor by the bound muon as a function of the incident center-of-mass energy. The arrow in the figure indicates the point where the total energy is zero.

Figure 2: Surface of section for 2 events, one has a small enhancement factor (top panels) and the other has a large enhancement factor (bottom panels), on the $x - p_x$ (left panels) and the $z - p_z$ (right panels) planes at the incident c.o.m energy 0.18keV, in the atomic unit.

Figure 3: Incident energy dependence of the sticking probability of the muon on the alpha particle. The statistical error is shown by error-bars, otherwise it is within the size of the points in the figure. (Left panel) Distance between the muon and the alpha particle as a function of the inter-nuclear separation. (Right panel)

Figure 4: Time $T$ map of a stuck muon on the *r-p* plane (left panels). One with the external oscillating force (right panels) with frequency $2\gamma_\mu$, both in the atomic unit.

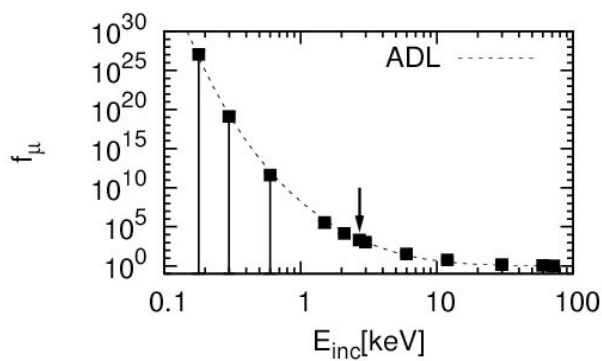

figure 1

figure 2

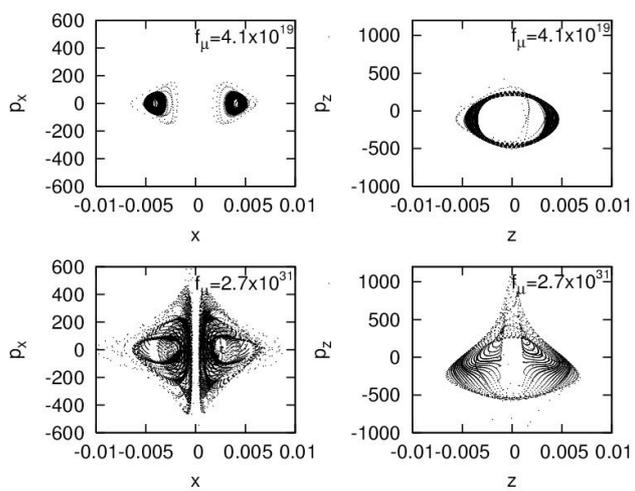

Figure 3

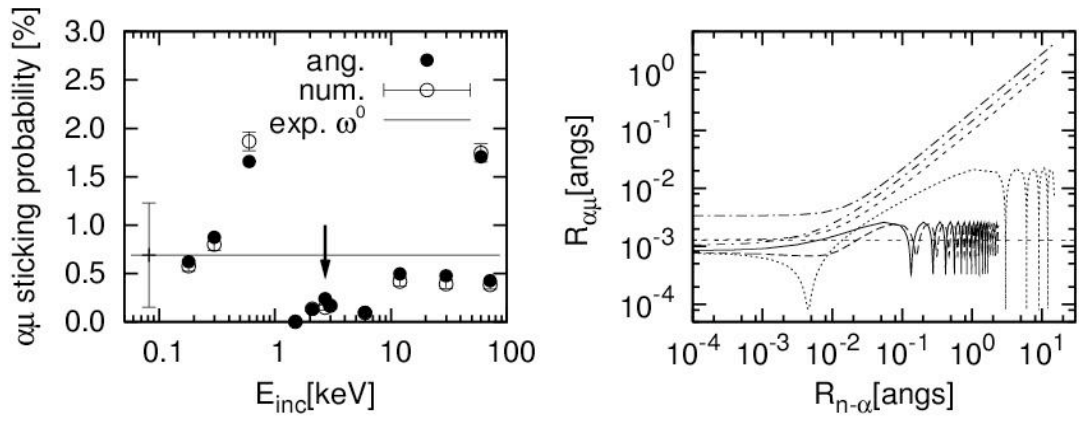

Figure 4

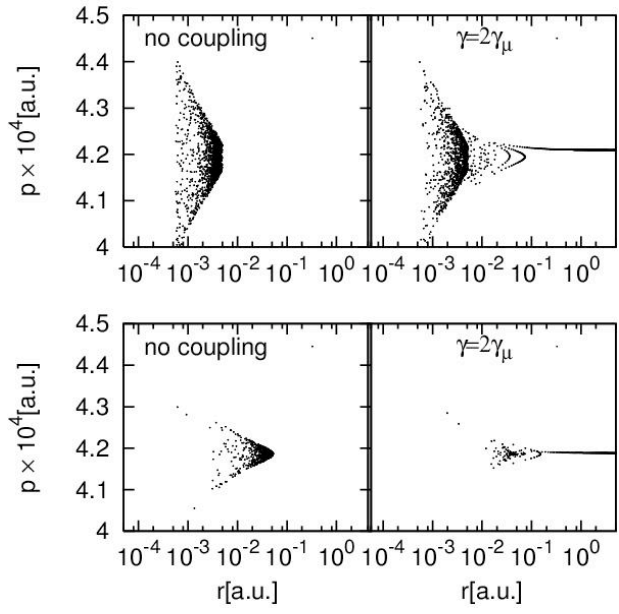